\documentclass[aps,prc,groupedaddress,twocolumn,showpacs,amsmath,amssymb]{revtex4}
\usepackage{graphics}
\usepackage{dcolumn}
\usepackage{longtable}

\begin{document}

\title{Further explorations of the $\alpha$-particle optical model potential at low energies for the mass range $A$$\approx$45--209}

\author{V.~Avrigeanu}
\email{vlad.avrigeanu@nipne.ro}
\author{M.~Avrigeanu}
\author{C.~M\u an\u ailescu}

\affiliation{Horia Hulubei National Institute for Physics and Nuclear Engineering, P.O. Box MG-6, 077125 Bucharest-Magurele, Romania}

\begin{abstract}
The recent high-precision measurements of $\alpha$-particle induced reaction data below the Coulomb barrier ($B$) make possible the understanding of limits and possible improvement of a previous optical model potential (OMP) for $\alpha$-particles on nuclei within the mass number range 45$\leq$$A$$\leq$209 [M. Avrigeanu {\it et al.}, Phys. Rev. C {\bf 82}, 014606 (2010)]. An updated version of this potential is given in the present work concerning mainly an increased surface imaginary potential depth well below $B$ for $A$$>$130. Moreover, underestimation of reaction cross sections for well-deformed nuclei is removed by using $\sim$7\% larger radius for the surface imaginary part of this spherical OMP. Improved input parameters based on recent independent data, particularly $\gamma$-ray strength functions,  but no empirical rescaling factor of the $\gamma$ and/or neutron widths have been involved within statistical model calculation of the corresponding $(\alpha,x)$ reaction cross sections. 
\end{abstract}


\maketitle

\section{INTRODUCTION}

The recent accurate $\alpha$-particle elastic-scattering and -induced reaction data around the Coulomb barrier ($B$) on heavier mass nuclei \cite{ap12,ggk11,ggk13,df11,ap120,ggk12,zh12,as11,gg10,jg12,ggk14,ggk11b,tr12,ln13} make possible the improvement of a previous $\alpha$-particle global optical model potential (OMP) \cite{ma09,ma10}. 
Thus, while the use of this potential led to description of some new data almost as well as local potentials \cite{ggk13,df11,ap120,ggk12,zh12,gg10,va11} or showing the best agreement of global potentials \cite{ap12,ggk11,pm11,pm13,tr11}, underestimation of newer data \cite{as11,jg12,ggk14,ggk11b,tr12,ln13} was also found. 
Hence a revision of this OMP \cite{ma10} became desirable and forms the object of the present work. It is also motivated by actual needs \cite{jg12,ggk14,ggk11b,tr12,tr11,sjq14} to obtain a full understanding of $\alpha$-induced reaction cross sections using spherical optical potential within statistical model of nuclear reactions. Suitable knowledge of this issue is also a condition for reanalysis of the $\alpha$-emission underestimation by OMPs that describe the $\alpha$-particle elastic scattering (\cite{ma06} and Refs. therein). 

The previous work on consistent description of $(\alpha,x)$ reactions \cite{ma10} followed up several earlier steps. First, we looked for the avoidance of the question marks related to (i) the rest of model parameters that are used to describe the compound-nucleus (CN) de-excitation through $\alpha$-particle emission (see, e.g., shaded areas in Figs. 4--9 of Ref. \cite{go02}), as well as (ii) the differences between the $\alpha$-particles in the incoming and outgoing channels. Thus, formerly we carried out an analysis of only elastic-scattering angular distributions of $\alpha$-particles on $A$$\sim$100 nuclei at energies below 35 MeV \cite{ma03}. A semi-microscopic OMP with a double-folding model (DFM) including the explicit treatment of the exchange component was used in this respect. 
A dispersive correction to the microscopic DFM real potential was also considered together with a phenomenological energy-dependent imaginary part that was finally obtained. Second, a full phenomenological analysis of the same data provided a regional optical potential (ROP), to be used in further nuclear-reaction model calculations. Next, similar semi-microscopic and phenomenological analyses concerned $A$$\sim$50--120 nuclei and energies from $\sim$13 to 50 MeV, but including furthermore an ultimate statistical-model (SM) assessment of the  available $(\alpha,\gamma)$, $(\alpha,n)$ and $(\alpha,p)$ reaction cross sections for target nuclei from $^{45}$Sc to $^{118}$Sn and incident energies below 12 MeV \cite{ma09,ma10b}. 
Third, the extension of the same analysis to heavy nuclei \cite{ma09b,ma10,va11} proved the essential role of the energy dependence of $\alpha$-particle surface imaginary potential depth below $B$. 

Results corresponding to the OMP of Ref. \cite{ma10} are compared in the present work with the $(\alpha,x)$ reaction data published in the meantime for heavier target nuclei. Improved SM input parameters which are based on recent independent data, particularly the $\gamma$-ray strength functions, for statistical model calculation of the corresponding $(\alpha,x)$ reaction cross sections are discussed in Sec.~\ref{MScalc}. A consequent OMP update is presented in Sec.~\ref{OMup}, including a particular adjustment for the well-deformed nuclei with 152$<$$A$$<$190. The results are discussed in Sec.~\ref{Disc}, followed by conclusions in Sec.~\ref{Conc}, while preliminary results were presented elsewhere \cite{ND2013,cssp14}.

\section{$(\alpha,x)$ reaction data analysis} \label{MScalc}

\begin{figure} [t]
\resizebox{0.8\columnwidth}{!}{\includegraphics{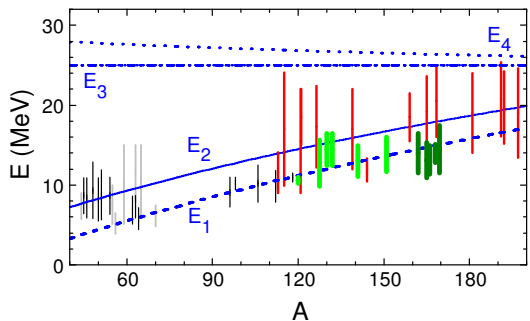}}
\caption{\label{Fig:E14}(Color online) The $A$-dependence of energies $E_1$ (dashed curve) below which the depth $W_D$ is constant, $E_2$ (solid curve) corresponding to 0.9$B$, $E_3$ (dash-dotted curve) and $E_4$ (dotted curve) given in Table~\ref{tab:newomp}, and the energy ranges of the $(\alpha,x)$ reaction data involved in this work for $A$$\geq$120 \cite{ap120,ggk12,zh12,as11,gg10,jg12,ggk14,ggk11b,tr12,ln13} (thick bars) as well as formerly for $A$$<$120 \cite{ma09} (thin bars), and $A$$>$113 \cite{ma10} (medium bars).}
\end{figure}

The more recent $(\alpha,x)$ reaction data concern mainly heavier target nuclei and incident energies well below $B$ \cite{df11,ap120,ggk12,zh12,as11,gg10,jg12,ggk14,ggk11b,tr12,ln13} (Fig.~\ref{Fig:E14}). They are partly supporting this potential and partly pointing out the need for an update. Thus, the new measurements for the $(\alpha,n)$ reaction on the lighter target nuclei $^{120}$Te \cite{ap120}, $^{127}$I \cite{ggk12}, $^{130,132}$Ba \cite{zh12}, and $^{151}$Eu \cite{gg10} as well as for the $(\alpha,\gamma)$ reaction on $^{127}$I and $^{151}$Eu  are rather well described by this potential. 
However, we have met difficulties in describing even the $(\alpha,n)$ reaction data for the heavier nucleus $^{141}$Pr \cite{as11} and especially the well-deformed nuclei $^{165}$Ho and $^{166}$Er \cite{jg12}, $^{162}$Er \cite{ggk14}, $^{169}$Tm \cite{ggk11b,tr12}, and $^{168}$Yb \cite{ln13}, as well as the $(\alpha,\gamma)$ reaction cross sections for $^{130}$Ba \cite{zh12} and well-deformed nuclei \cite{ggk14,ggk11b,tr12,ln13}. Consequently, further efforts had to be devoted to the OMP parameters for heavier nuclei and the distinct case of the deformed nuclei, as well as to the account of the $\gamma$-ray strength functions,  to clarify the role of the $\alpha$-particle OMP within the later cases.
However, a discussion should concern firstly the possible role of the Coulomb excitation (CE) process in the establishment of the $\alpha$-particle OMP through the $(\alpha,x)$ reaction data analysis.

\subsection{Coulomb excitation effects on $\alpha$-particle OMP setting up} 

The possible effects of the CE consideration on the so-called "$\alpha$-potential mystery" have been recently underlined by Rauscher \cite{tr12b,tr13}. First, it has been pointed out that even the numerical methods employed to determine the Coulomb wave functions for low energy and high Coulomb barrier play an important role within $(\alpha,x)$ reaction data analysis. Thus it was shown that a large difference exists between the results obtained for the $^{144}$Sm$(\alpha,\gamma)^{148}$Gd reaction cross sections using either the old routine \cite{tr01} for Coulomb transmission, or a new one \cite{tr12b}. Concerning the calculations carried on using the code SCAT2 \cite{scat2}, which was involved earlier \cite{ma09,ma10,ma06,ma03} as well as within present work, their correctness in this respect is proved directly by the corresponding results shown formerly \cite{ma10} and in Fig. 1 of Ref. \cite{tr12b} as being obtained with the new routine and the OMP of Ref. \cite{ma10}.

Second, the CE has been considered as an additional reaction channel which is competing, at the $\alpha$-particle energies well below $B$, with the CN formation while it is not present within $\alpha$-particle emission from an excited CN \cite{tr12b,tr13}. Because CE was not considered in the worldwide used optical potential \cite{mcf66}, which was obtained by $\alpha$-particles elastic-scattering analysis and then used in calculation of $\alpha$-particle emission data, Rauscher \cite{tr12b,tr13} has adopted a decreased CN formation cross section for the $\alpha$-induced reactions. This reduction of the CN formation cross section given by the OMP has been obtained by taking into account, for each partial wave, the CE cross section that should be additionally considered for that partial wave. Next, a further reduction by a factor of 3 was found necessary  to describe the measured $^{144}$Sm$(\alpha,\gamma)^{148}$Gd reaction cross sections \cite{es98}. Finally, it was shown that this approach is necessary for the description of the above-mentioned $(\alpha,\gamma)$ reaction and the $(\alpha,n)$ reaction on the target nuclei $^{141}$Pr and $^{169}$Tm, but not for the $(\alpha,\gamma)$ reaction on $^{168}$Yb and $(\alpha,n)$ reaction on $^{130,132}$Ba and $^{168}$Yb \cite{tr12b,tr13}, and subsequently for both reactions on $^{113}$In \cite{ggk13}.

While the decrease of the total reaction cross section $\sigma_R$ owing to the direct-interaction channels is usually taken into account in SM calculations of reaction cross sections, the case of CE is indeed quite different. A reference paper in this respect was given, however, by Vonach {\it et al.} \cite{hv83} on $\alpha$-particle $\sigma_R$ derived from $(\alpha,n)$ reaction cross sections through extensive SM calculations. They pointed out that, because the CE cross section becomes the dominant part of the nonelastic cross section below the Coulomb barrier, their results obtained on the basis of the measured $(\alpha,n)$ reaction cross sections and SM calculations do not represent indeed the full nonelastic cross section but "they do, however, correctly describe the CN formation cross section needed in statistical model calculations". 
The use of the notation of $\sigma_R$ for these SM results, even under these conditions, may also have a theoretical support. Thus, Hussein {\it et al.} \cite{msh91} showed that the derivation of $\sigma_R$ through the use of optical theorem has the same results either paying no attention to the long-range Coulomb interaction, or including the Rutherford scattering amplitude within a generalized approach.
However, a decomposition of $\sigma_R$ into a direct reaction  contribution and a fusion cross section has a straightforward schematic representation in terms of partial waves, and quite distinct from the elastic-scattering and CE cross sections, only for heavy-ion interactions, and under semiclassical conditions (e.g., Fig. 1.8 of Ref. \cite{wn80}). Therefore, we should note that the approach of  Vonach {\it et al.} \cite{hv83} will be followed in the present work, with the understanding that the quantity $\sigma_R$ corresponds to the CN formation. Additional coupled-channels (CC) calculations, which are obviously handling CE processes, are beyond the object of the OMP \cite{ma10} present revision.

\subsection{Statistical model parameters}

\begingroup
\squeezetable
\begin{table*} 
\caption{\label{densp} Low-lying levels number $N_d$ up to excitation energy $E^*_d$ \protect\cite{ensdf,xundl} used in cross-section calculations, and the levels, $s$-wave neutron-resonance spacings $D_0^{\it exp}$ and average radiation widths $\Gamma_{\gamma}$ in the energy range $\Delta$$E$ above the separation energy $S$ (with uncertainties given in parentheses, in units of the last digit)  \cite{ripl3}, for the target-nucleus ground-state spin $I_0$, fitted to obtain the BSFG level-density parameter {\it a} and ground-state shift $\Delta$ (for a spin cutoff factor calculated with a variable moment of inertia \cite{va02} between half and 75\% of the rigid-body value, from ground state to $S$, and reduced radius $r_0$=1.25 fm), and the EGLO model parameters $k_0$ and $T_f$ corresponding to description of the RSF data \cite{ripl2,mg14} and $\Gamma_{\gamma}$ values.}
\begin{ruledtabular}
\begin{tabular}{ccccccccccrcc}\\
Nucleus &$N_d$&$E^*_d$& \multicolumn{6}{c}
  {Fitted level and resonance data}& $a$ & $\Delta$\hspace*{3mm}& $k_0$ & $T_f$\\
\cline{4-9}
  &  &    &$N_d$&$E^*_d$&$S+\frac{\Delta E}{2}$&$I_0$& $D_0^{\it exp}$ & $\Gamma_{\gamma}$ \\ 
          &  &(MeV)&  &(MeV)& (MeV) &   &   (keV)  & (meV) &(MeV$^{-1}$)&(MeV)& & (MeV)\\ 
\hline
$^{113}$In&33&1.768&33&1.768&       &   &          &       &14.20& 0.00 \\ 
$^{116}$Sn&17&2.844&26&3.106&       &   &          &       &13.45& 1.33 \\ 
$^{116}$Sb&22&0.881&22&0.881&       &   &          &       &14.60&-0.70 \\
$^{117}$Sb&17&1.536&17&1.536&       &   &          &       &14.10& 0.05 \\  

$^{120}$Te&20&2.461&20&2.461&       &   &          &       &14.00& 0.87 \\ 
$^{123}$ I&31&1.453&31&1.453&       &   &          &       &14.00&-0.35 \\ 
$^{123}$Xe&31&0.876&31&0.876&       &   &          &       &14.50&-0.87 \\ 
$^{124}$Xe&31&2.382&29&2.373&       &   &          &       &14.20& 0.64 \\ 

$^{127}$ I&33&1.480&33&1.480&       &   &          &       &14.00&-0.35 \\ 
$^{130}$Xe&26&2.442&26&2.442& 9.256 &1/2& 0.038(5) &       &13.80& 0.68 \\ 
$^{130}$Cs&11&0.318&11&0.318&       &   &          &       &14.00&-1.32 \\ 
$^{131}$Cs&20&1.048&23&1.212&       &   &          &       &13.60&-0.54 \\ 

$^{130}$Ba&19&2.101&25&2.280&       &   &          &       &14.00& 0.57 \\
$^{132}$Ba&32&2.505&25&2.374&       &   &          &       &13.80& 0.63 \\  
$^{133}$La&28&1.319&28&1.319&       &   &          &       &14.00&-0.50 \\
$^{135}$La&17&1.038&17&1.038&       &   &          &       &13.50&-0.60 \\ 
$^{133}$Ce&21&1.201&21&1.201&       &   &          &       &14.00&-0.45 \\ 
$^{134}$Ce&16&2.050&24&2.304&       &   &          &       &13.80& 0.56 \\ 
$^{135}$Ce&13&1.367&13&1.367&       &   &          &       &13.80&-0.10 \\ 
$^{136}$Ce&14&2.451&14&2.451&       &   &          &       &13.20& 0.88 \\ 

$^{141}$Pr&23&1.853&45&2.190&       &   &          &       &13.50& 0.10 \\ 
$^{144}$Nd&52&2.779&52&2.779& 7.917 & 0 & 0.038(2) &       &15.00& 0.84 \\ 
$^{144}$Pm&11&0.363&11&0.363&       &   &          &       &15.50&-1.02 \\ 
$^{145}$Pm&28&1.397&28&1.397&       &   &          &       &15.50&-0.21 \\ 

$^{144}$Sm&21&2.883&21&2.883&       &   &          &       &15.00& 1.34 \\
$^{148}$Sm&33&2.228&32&2.214& 8.141 &7/2& 0.0057(5)& 69(4) &17.00& 0.70 &  1  & 0.45\\
$^{149}$Sm&22&0.881&22&0.881& 5.871 & 0 & 0.065(20)& 44(4) &17.68&-0.44 & 1.3 & 0.45\\ 
$^{147}$Eu&28&1.421&18&1.244&       &   &          &       &17.30&-0.03 \\
$^{151}$Eu&30&0.654&30&0.654&       &   &          &       &17.00&-0.88 \\  
$^{147}$Gd&15&1.701&15&1.701&       &   &          &       &17.30& 0.49 \\ 
$^{148}$Gd&23&2.700&20&2.633&       &   &          &       &17.00& 1.30 \\
$^{156}$Gd&25&1.540&25&1.540& 8.536 &3/2& 0.0017(2)&108(10)&18.00& 0.20 &  3  & 0.3 \\ 
$^{157}$Gd&50&0.840&54&0.888& 6.360 & 0 & 0.030(6) & 88(12)&17.90&-0.76 &  3  & 0.33\\
$^{158}$Gd&25&1.452&25&1.452& 7.937 &3/2& 0.0049(5)& 97(10)&17.35& 0.06 &  3  & 0.3 \\  

$^{155}$Tb&22&0.616&22&0.616&       &   &          &       &17.50&-0.77 \\
$^{156}$Tb&18&0.405&15&0.313&       &   &          &       &18.15&-0.95 \\

$^{160}$Dy&39&1.676&39&1.676&       &   &          &       &17.50& 0.13 &  2  & 0.3 \\
$^{161}$Dy&32&0.641&26&0.568& 6.455 & 0 & 0.027(5) &108(10)&17.80&-0.86 &  2  & 0.3 \\
$^{162}$Dy&27&1.575&27&1.575& 8.197 &5/2& 0.0024(2)&112(10)&17.64& 0.17 &  2  & 0.3 \\
$^{163}$Dy&33&0.740&34&0.766& 6.271 & 0 & 0.062(5) &112(20)&17.20&-0.80 &  2  & 0.3 \\
$^{164}$Dy&18&1.346&18&1.346& 7.658 &5/2& 0.0068(6)&113(13)&16.92& 0.02 &  2  & 0.3 \\
$^{165}$Ho&24&0.744&24&0.744&       &   &          &       &18.00&-0.62 \\

$^{162}$Er&21&1.506&21&1.506&       &   &          &       &17.30& 0.16 \\
$^{166}$Er&27&1.760&27&1.760&       &   &          &       &17.20& 0.30 &  2  & 0.31\\
$^{167}$Er&36&0.813&39&0.856& 6.436 & 0 & 0.038(3) & 92(8) &17.91&-0.69 &  2  & 0.36\\  
$^{168}$Er&25&1.493&21&1.422& 7.772 &7/2& 0.0042(3)&       &17.20& 0.05 \\ 
$^{168}$Tm&20&0.245&27&0.366&       &   &          &       &18.35&-1.12 \\ 
$^{169}$Tm&21&0.646&21&0.646&       &   &          &       &18.20&-0.67 \\ 

$^{165}$Yb&24&0.670&23&0.665&       &   &          &       &17.30&-0.76 \\ 
$^{166}$Yb&27&1.617&27&1.617&       &   &          &       &17.50& 0.19 \\ 
$^{168}$Yb&28&1.551&28&1.551&       &   &          &       &17.50& 0.11 \\ 
$^{169}$Yb&30&0.762&29&0.758& 6.867 & 0 & 0.008(3) &       &19.20&-0.58 \\ 
$^{170}$Yb&26&1.521&37&1.669& 8.470 &7/2& 0.0016(4)& 80(25)&17.50& 0.11 &  2  & 0.25\\ 
$^{171}$Yb&20&0.780&40&1.004& 6.615 & 0 & 0.0035(6)& 70(10)&18.10&-0.52 &  2  & 0.27\\
$^{172}$Yb&22&1.510&41&1.720& 8.020 &1/2& 0.0069(5)& 75(5) &18.20& 0.20 &  2  & 0.28\\
$^{171}$Lu&26&0.671&26&0.671&       &   &          &       &18.10&-0.73 \\  
$^{172}$Lu&28&0.406&28&0.406&       &   &          &       &18.50&-1.05 \\ 
$^{173}$Lu&26&0.735&26&0.735&       &   &          &       &18.35&-0.64 \\ 
$^{171}$Hf&23&0.867&23&0.867&       &   &          &       &17.50&-0.53 \\ 
$^{172}$Hf&26&1.534&26&1.534&       &   &          &       &18.20& 0.16 \\ 
\end{tabular}	 
\end{ruledtabular}
\end{table*}
\endgroup

We have also used within actual $(\alpha,x)$ reaction analysis a consistent set of nucleon \cite{KD03} and $\gamma$-ray transmission coefficients, and back-shifted Fermi gas (BSFG) nuclear level densities \cite{hv88}. They have been established or validated on the basis of independent experimental information for neutron total cross sections \cite{exfor}, $\gamma$-ray strength functions, and low-lying levels \cite{ensdf,xundl} and resonance data \cite{ripl3}, respectively. 
Hereafter only the points in addition to the details given formerly \cite{ma09,ma10} are mentioned as well as the particular parameter values that could be used within further trials, while the SM calculations were carried out using an updated version of the computer code STAPRE-H95 \cite{ma95}.
Thus, the BSFG parameters used in the following as well as the independent data used for their setting up are given in Table~\ref{densp}.
However, because of the difficulties found in describing the above-mentioned $(\alpha,\gamma)$ reaction cross sections, an additional effort was devoted to the account of $\gamma$-ray strength functions, unlike the renormalization \cite{ggk13} of default $\gamma$-ray transmission coefficients to achieve agreement with the $(\alpha,\gamma)$ data. We took in this respect the opportunity of high accuracy measurements of the radiative strength function (RSF) performed within the latest years especially at lower energies through the well-known Oslo method \cite{as00,mg14}, leading to the RSF models progress. 

\begin{figure} [t]
\resizebox{1.0\columnwidth}{!}{\includegraphics{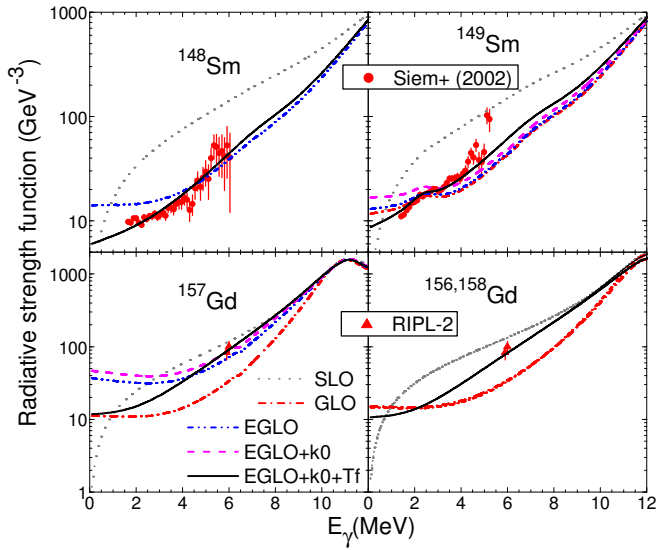}}
\caption{\label{Fig:RSF-NbGd3}(Color online) Comparison of measured \cite{ripl2,mg14} and calculated electric-dipole $\gamma$-ray strength functions for the $^{148,149}$Sm, and $^{156-158}$Gd nuclei, using the SLO (dotted curves), GLO (dash-dotted curves), and EGLO (dash-dot-dotted curves) models, including the effects of using the free parameters $k_0$ (dashed curves) and $T_f$ (solid curves) given in Table~\ref{densp}. In the case of two isotopes of the same element, the shorter or thinner curve corresponds to the latter one. The measured \cite{ripl3} $s$-wave neutron-resonance average radiation widths $\Gamma_{\gamma}$  are given in Table~\ref{densp} while the calculated values using each model can be found elsewhere \cite{cssp14}.}
\end{figure}

\begin{figure} [t]
\resizebox{1.0\columnwidth}{!}{\includegraphics{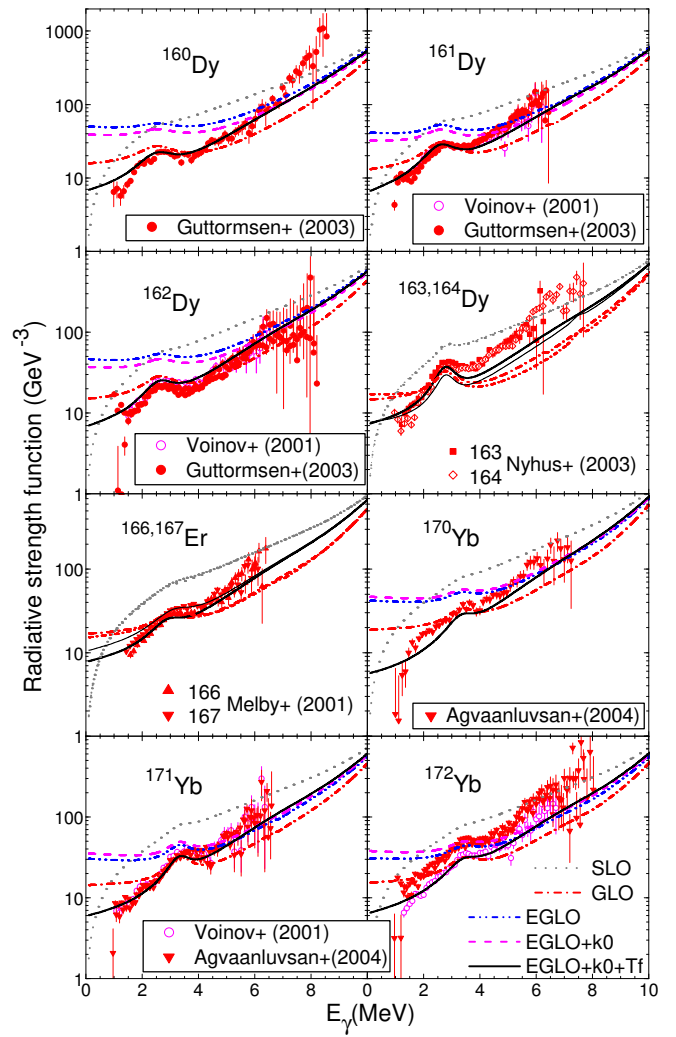}}
\caption{\label{Fig:RSF-DyErYb}(Color online) As Fig.~\ref{Fig:RSF-NbGd3} but for the sum of $\gamma$-ray strength functions of the $E1$ and $M1$ radiations for the $^{160-164}$Dy, $^{166,167}$Er, and $^{170-172}$Yb nuclei.}
\end{figure}

The former Lorentzian (SLO) model for the electric-dipole $\gamma$-ray strength functions, of main importance for calculation of the $\gamma$-ray  transmission coefficients, has used the giant dipole resonance (GDR) line shape with the usual parameters ($\sigma_0, \Gamma_0$, and $E_0$) derived from photoabsorption data (\cite{ripl3} and Refs. therein).
Later, an energy dependence of the GDR width $\Gamma(E_{\gamma})$ was assumed also within the energy-dependent Breit-Wigner (EDBW) model \cite{dgg79,ma87} that was formerly involved \cite{ma09,ma10}. The generalized Lorentzian (GLO) model of Kopecky and Uhl \cite{jk91} has included in addition a further dependence on the nuclear temperature $T_f$ of the final states, to avoid the extrapolation of the SLO function in the limit of zero $\gamma$-ray energy but a rather constant nonzero limit. 
Moreover, the enhanced generalized Lorentzian (EGLO) model \cite{jk93,ripl3} assumes also an enhancement of the GLO width $\Gamma(E_{\gamma},T_f)$, going from $k_0$ at a $\gamma$-ray energy $\epsilon_0$ to unity at $E_0$,
\begin{equation}\label{eq:2}
\Gamma(E_{\gamma},T_f)=\left[k_0+\frac{E_{\gamma}-\epsilon_0}{E_0-\epsilon_0}(1-k_0)\right] \frac{\Gamma_0}{E_0^2}(E_{\gamma}^2+4\pi^2T_f^2)    \:\: ,
\end{equation} 
with the values of the two parameters $k_0$ and $\epsilon_0$=4.5 MeV adjusted to reproduce the averaged resonance capture data. However, we found differences between the $k_0$ values given by the latest RIPL-3 form of Eqs. (143) \cite{ripl3} and (6.9) \cite{ripl1}, and the related content in Fig. 6.1 of RIPL-1 \cite{ripl1} (e.g., 2.49 and 2.00, respectively, for a nucleus with $A$=158). 
At the same time we took into account the recent analysis \cite{bb13} of the effects owing to the assumption of the temperature $T_f$ variation from zero to the value corresponding to the BSFG model. Consequently, following also \cite{acl10} and Refs. therein, we have looked for both $k_0$ and $T_f$ constant values that correspond to description of the RSF data \cite{ripl2,mg14} and $s$-wave neutron-resonance average radiation widths $\Gamma_{\gamma}$ \cite{ripl3} for the heavier nuclei of interest for the present work (Table~\ref{densp}). 

The effects of the $k_0$ and $T_f$ values on the RSF calculation using the EGLO model, along with the corresponding results provided by the SLO and GLO models, are shown in Figs.~\ref{Fig:RSF-NbGd3} and \ref{Fig:RSF-DyErYb}. The GDR as well as pigmy dipole resonance parameters established within the original references \cite{ripl2,mg14} have been used in this respect.
Concerning the M1 radiation, the above-mentioned SLO model was used along with either the global parametrization  \cite{ripl3} for the GDR energy and width,  i.e. $E_0$=41$\cdot$$A^{1/3}$ MeV and $\Gamma_0$=4 MeV, and the value of $\sigma_0$ derived from the systematics of $f_{M1}(E_{\gamma}$=7 MeV)=1.58$\cdot10^{-9}A^{0.47\pm0.21}$ (Eq. (6.12) of Ref. \cite{ripl1}), or particular GDR parameters.
The $\Gamma_{\gamma}$ values for nuclei without resonance data have been estimated from systematics of the available data plotted against the neutron-separation energy for the even-even as well as odd isotopes (e.g. \cite{hkt11}). 

Further discussion of the sensitivity of calculated $(\alpha,\gamma)$ reaction cross sections to the adopted $f_{E1}(E_{\gamma})$ model is given below for the particularly questionable case of the $^{168}$Yb target nucleus.

\section{Updated OMP} \label{OMup}
\subsection{Updated surface imaginary potential depth}

\begin{figure}  [t]
\resizebox{1.0\columnwidth}{!}{\includegraphics{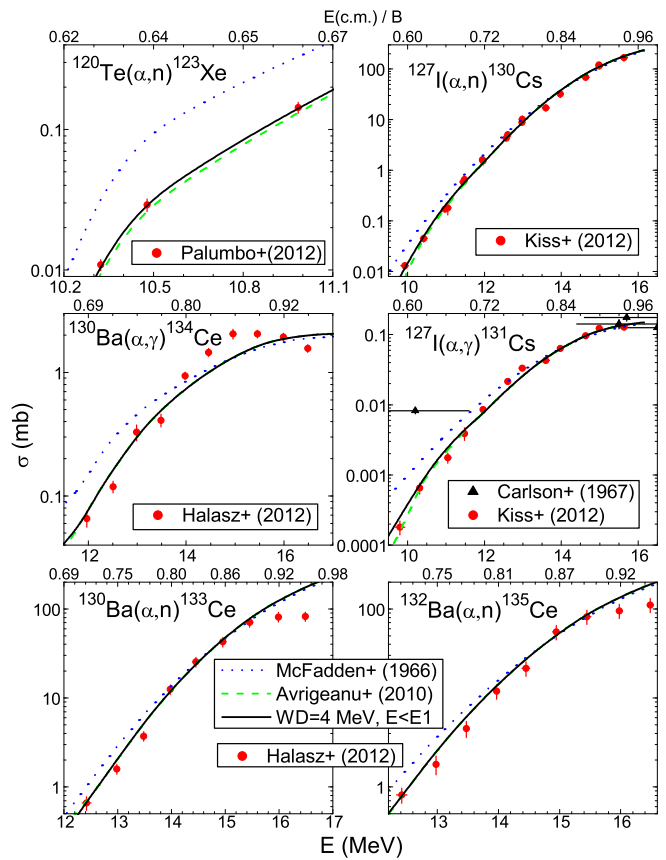}}
\caption{\label{Fig:TeIBa5} (Color online) Comparison of former \cite{exfor} and recently measured $(\alpha,x)$ reaction cross sections, for the target nuclei $^{120}$Te \cite{ap120}, $^{127}$I \cite{ggk12}, and $^{130,132}$Ba \cite{zh12}, and SM-calculated values using the $\alpha$-particle OMPs of Refs. \cite{mcf66} (dotted curves), \cite{ma10} (dashed curves), and Table~\ref{tab:newomp} of this work (solid curves).}
\end{figure}
\begin{figure}  [h]
\resizebox{1.0\columnwidth}{!}{\includegraphics{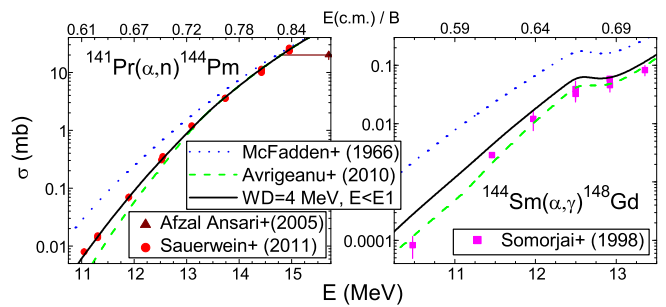}}
\caption{\label{Fig:PrSm} (Color online) As Fig.~\ref{Fig:TeIBa5} but for the target nuclei $^{141}$Pr \cite{as11} and $^{144}$Sm \cite{es98}.}
\end{figure}

The main attribute of the recently measured cross sections of $(\alpha,x)$ reactions on heavier nuclei \cite{ap120,ggk12,zh12,as11,gg10,jg12,ggk14,ggk11b,tr12,ln13} is their focus on energies below $B$, unlike the data previously available \cite{ma10}. 
A first group of data consists of the $(\alpha,x)$ reaction cross sections for $^{120}$Te \cite{ap120}, $^{127}$I \cite{ggk12}, $^{130,132}$Ba \cite{zh12}, and $^{151}$Eu \cite{gg10} target nuclei. 
SM calculations carried out using the global potential \cite{ma10} are compared to them in Fig.~\ref{Fig:TeIBa5} for the lighter nuclei while the former results obtained for $^{151}$Eu \cite{va11} are not significantly changed. 
The rather good agreement found for these reactions is, however, first attributable to either the target $A$$<$130 \cite{ap120,ggk12} or the corresponding energy ranges \cite{zh12,gg10} mainly above the energy limit $E_1$.
A particular case is, however, the $(\alpha,n)$ reaction on $^{120}$Te \cite{ap120} within an energy range which is fully below this energy limit $E_1$. Thus, it makes possible a suitable assessment of the value of surface imaginary potential depth $W_D$=3.5 MeV  for $A$$<$130 (see the note $b$ in Table I of Ref. \cite{ma10}). However, an even better description, beyond the former one within the error bars of these quite accurate data, is provided by a slightly increased value $W_D$=4 MeV along with the rest of the same OMP parameters given again in Table~\ref{tab:newomp}. 

\begin{table*} [t]
\caption{\label{tab:newomp}$\alpha$-particle OMP parameters (within the formalism of, e.g., Ref. \cite{KD03}) for target nuclei with 45$\leq$$A$$\leq$209 at energies $E$$<$50 MeV, in addition to the Coulomb potential of a uniformly charged sphere of reduced radius $r_C$=1.3 fm. The energies and range limits$^a$ are in MeV. A star used as superscript follows the parameters which were changed with respect to Ref. \cite{ma10}, while uncertainties of the mass-, charge-, and energy-dependence factors of the OMP parameters are given under these factors.}
\begin{tabular}{lclc}
\hline \hline
   \hspace*{0.9in}Potential depth (MeV)&               & \hspace*{0.9in}Geometry parameters (fm)\\ 
\hline
V$_R$=165 + 0.733 Z/A$^{1/3}$ - 2.64 E,                & E$\leq$E$_3$ & r$_R$=1.18 + 0.012 E,    & E$\leq$25\\
   \hspace*{0.25in}($\pm$6) ($\pm$0.094) \hspace*{0.30in}($\pm$0.17)& &\hspace*{0.15in}($\pm$0.05) ($\pm$0.002)\\
   \hspace*{0.14in}
     =116.5 + 0.337 Z/A$^{1/3}$ - 0.453 E,             & E$>$E$_3$       &\hspace*{0.15in}=1.48,& E$>$25\\
   \hspace*{0.26in}($\pm$4.6) ($\pm$0.101) \hspace*{0.32in}($\pm$0.112)& &\hspace*{0.15in}($\pm$0.04)\\ 
                                                       &               &
                                                 a$_R$=0.631 + (0.016 - 0.001E$_2$) Z/A$^{1/3}$,&E$\leq$E$_2$\\
										                          & &\hspace*{0.15in}($\pm$0.016) ($\pm$0.002)\\ 
                                                       &               &
                      \hspace*{0.15in}=0.631 + 0.016 Z/A$^{1/3}$ - (0.001 Z/A$^{1/3}$)E,&E$_2$$<$E$\leq$E$_4$\\
								  & & \hspace*{0.15in}($\pm$0.016) ($\pm$0.002)\\  
                                                       &               &
                      \hspace*{0.15in}=0.684 - 0.016 Z/A$^{1/3}$ - (0.0026-0.00026 Z/A$^{1/3}$)E,&E$>$E$_4$\\
									& & \hspace*{0.15in}($\pm$0.016) ($\pm$0.002)\\  
W$_V$=2.73 - 2.88 A$^{1/3}$ + 1.11 E                   &               & r$_V$=1.34\\
   \hspace*{0.25in}($\pm$3.5)($\pm$0.87) \hspace*{0.20in}($\pm$0.04) & & a$_V$=0.50\\ 
W$_D^*$=4 ,                                            & E$\leq$E$_1$  & r$_D^*$=1.52 ,\hspace*{1.02in} 152$\geq$A$\geq$190\\
  \hspace*{0.25in}(+0.5/-2.5)                          &               & \hspace*{0.15in}($\pm$0.04)\\ 
  \hspace*{0.22in}=	22.2 + 4.57 A$^{1/3}$ - 7.446 E$_2$+6E, & \: E$_1$$<$E$\leq$E$_2$ \:
	                                                     &\hspace*{0.15in}=max(1.74-0.01E,1.52) ,\hspace*{0.05in} 152$<$A$<$190\\ 
  \hspace*{0.26in}($\pm$4.4) ($\pm$0.98)             & &\hspace*{0.15in}($\pm$0.06)\\    
  \hspace*{0.22in}=	22.2 + 4.57 A$^{1/3}$ - 1.446 E ,& E$>$E$_2$       & a$_D$=0.729-0.074A$^{1/3}$\\
  \hspace*{0.26in}($\pm$4.4) ($\pm$0.98) \hspace*{0.23in}($\pm$0.08)\\     
\hline \hline
\multicolumn{4}{l}{$^a$E$_1^*$=-3.03-0.762A$^{1/3}$+1.24E$_2$; 
                       E$_2$=(2.59+10.4/A)Z/(2.66+1.36A$^{1/3}$); 
											 E$_3$=22.2+0.181Z/A$^{1/3}$; E$_4$=29.1-0.22Z/A$^{1/3}$}. \\
\end{tabular}
\end{table*}

\begin{figure}  [t]
\resizebox{1.0\columnwidth}{!}{\includegraphics{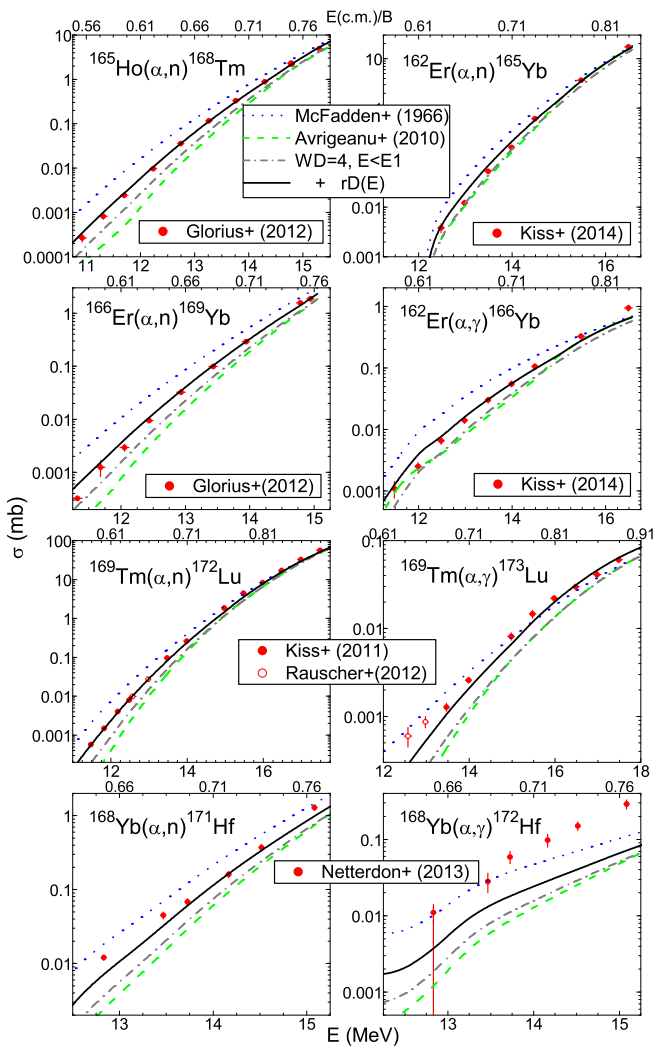}}
\caption{\label{Fig:HoErTmYb} (Color online) As Fig.~\ref{Fig:TeIBa5} but for $^{165}$Ho \cite{jg12}, $^{162}$Er \cite{ggk14}, $^{166}$Er \cite{jg12}, $^{169}$Tm \cite{ggk11b,tr12}, $^{168}$Yb \cite{ln13}, and SM calculations using also the OMP of this work without the $\sim$7\% larger radius for the surface imaginary potential (dash-dotted curves).}
\end{figure}

\begin{figure}  [t]
\resizebox{1.0\columnwidth}{!}{\includegraphics{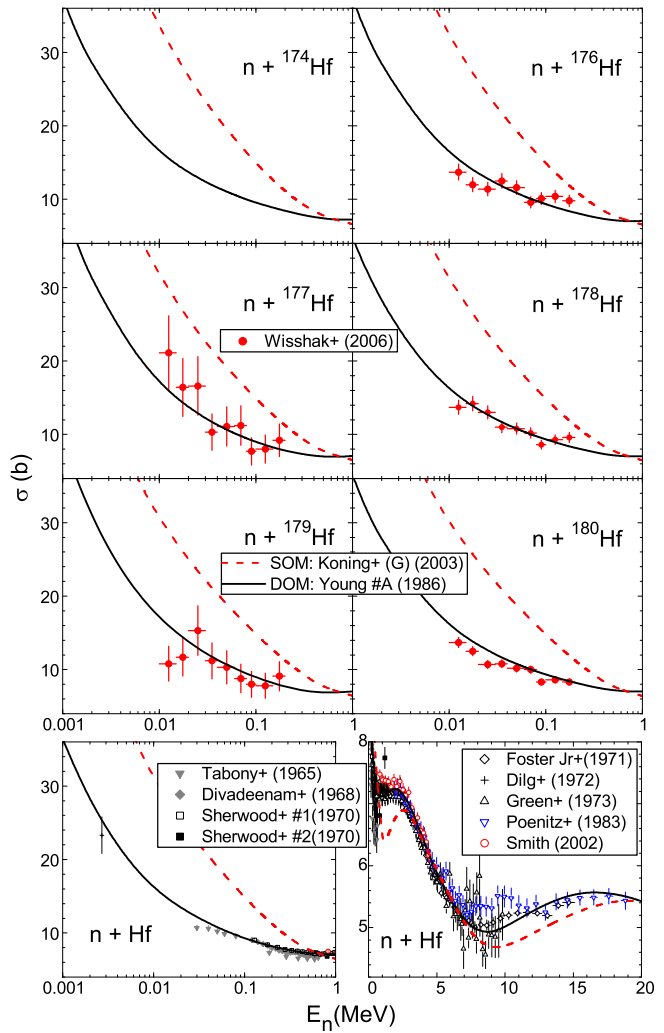}}
\caption{\label{Fig:HFntot} (Color online) Comparison of measured \cite{kw06,exfor} and calculated neutron total cross sections for Hf isotopes and the natural Hf by using the spherical \cite{KD03} (dashed curves) and deformed \cite{pgy86} (solid curves) OMPs. }
\end{figure}

However, the new quite accurate data for the $^{141}$Pr$(\alpha,n)^{144}$Pm reaction \cite{as11} have been more helpful in setting up the correct value of the $W_D$ parameter at the lowest energies. Thus, a significant underestimation of the data just below the energy limit $E_1$ by the OMP \cite{ma10} is entirely removed using the value $W_D$=4 MeV (Fig.~\ref{Fig:PrSm}). 

Alternatively, this value leads to an overestimation of the lowest-energy data points of the well-known $^{144}$Sm$(\alpha,\gamma)^{148}$Gd reaction data \cite{es98} that actually triggered the lower value $W_D$$\sim$1.5 MeV at energies $E$$\le$$E_1$ for $A$$>$130 \cite{ma10} owing to their uniqueness and incident-energy error bars as small as they are usual nowadays but rarely at the end of the '90s (see, e.g., Figs.~\ref{Fig:TeIBa5} and \ref{Fig:PrSm}).
All other data for heavier nuclei formerly analyzed (Figs. 4-5 of Ref. \cite{ma10}) are also better described by the increased $W_D$ parameter below $E_1$ that is shown in Table~\ref{tab:newomp} but with an uncertainty given by the above-mentioned difference. Because the rest of the OMP parameters are the average mass-, charge-, and energy-dependent values of the local parameters formerly obtained by analysis of $\alpha$-particle elastic-scattering angular distributions \cite{ma09,ma10}, the related standard deviations around the average values could be considered as uncertainties of the corresponding dependence factors. Former similar values were given for $A$$\sim$90 \cite{ma03} while more details may be found in Sec. 3 of Ref. \cite{ma09}. 

\subsection{Updated OMP for well-deformed nuclei}

A different case is that of the recent measured data for the well-deformed nuclei $^{165}$Ho, $^{162,166}$Er, $^{169}$Tm, and $^{168}$Yb \cite{jg12,ggk14,ggk11b,tr12,ln13}. Thus, the underestimation of both $(\alpha,n)$ and $(\alpha,\gamma)$ reactions is evident  for the OMP of Ref. \cite{ma10} as well as for its updated parameter value $W_D$=4 MeV (Fig.~\ref{Fig:HoErTmYb}). A particular note on the present reaction cross-section analysis for these nuclei should concern the use of the updated potential also for the calculation of the collective inelastic-scattering cross sections by means of the direct-interaction distorted-wave Born approximation (DWBA) method and recommended deformation parameters \cite{ripl3} within a local version of the computer code DWUCK4 \cite{pdk84}. Typical ratios of the direct inelastic scattering to the total reaction cross sections for, e.g., the $^{162}$Er target nucleus, from the threshold energy for the $(\alpha,n)$ reaction up to 16.5 MeV \cite{ggk14}, increase from $\sim$0.1\% to 7\%. They have been used for the corresponding $\sigma_R$ decrease within the rest of the CN calculations, with no real effect on the underestimation of the measured data.

The use of a spherical OMP \cite{KD03} in the neutron-emission channel instead of a deformed optical potential, so well motivated for the deformed nuclei, was the first issue deserving a careful analysis. Therefore we have replaced the former neutron transmission coefficients with the ones obtained by using the average rare-earth-actinide deformed phenomenological optical potential of Young \cite{pgy86} (Set A) within CC calculations, and deformation parameters given recently \cite{gn09} for Hf isotopes. The computer code EMPIRE-II \cite{mh02} was used in this respect.
First, we found that the measured neutron total cross sections \cite{kw06,exfor} are obviously described only by the deformed OMP at energies of tens of keV as well as between 1--3 and 7--20 MeV (Fig.~\ref{Fig:HFntot}). Second, the calculated $^{168}$Yb$(\alpha,x)$$^{171,172}$Hf reaction cross sections remained however unchanged after this replacement, mainly owing to the similar neutron total cross sections given by the two OMPs around the evaporation energy of $\sim$1 MeV. This conclusion has been quite useful also for the present analysis of neutron-emission in $\alpha$-particle induced reactions on well-deformed nuclei, carried out on the basis of the spherical neutron OMP \cite{KD03}.

Alternately, one should take into account the fact that nuclear deformation also motivates a low-energy enhancement of the charged-particle reaction cross sections as it was proved by Lanier {\it et al.} \cite{rgl90} for protons on $^{151,153}$Eu and recalled recently by Grimes \cite{smg13}. Thus, Lanier {\it et al.} pointed out that the enhancement of $(p,n)$ reaction cross sections for $^{153}$Eu relative to $^{151}$Eu, with large difference between the corresponding ground state deformations of these nuclei, can be accounted for if spherical OMP calculations are performed with an $\sim$3\% larger radius for $^{153}$Eu.

\begingroup
\squeezetable
\begin{table*} 
\caption{\label{tab:localomp} Optical potential parameters and volume integrals (without the negative sign) obtained by fit of the $\alpha$-particle elastic-scattering angular distributions of given Refs. at energies $\leq$50 MeV, for $A$$>$130 target nuclei.}
\begin{ruledtabular}
\begin{tabular}{ccccccccccc} 
Target &$E_{\alpha}$& Ref.& $V_R$ & $r_R$ & $a_R$ & $W_V$ & $W_D$ & $J_R$          & $J_V$           & $J_D$ \\
nucleus&    (MeV)   &     & (MeV) &  (fm) &  (fm) & (MeV) & (MeV) &MeV$\cdot$fm$^3$& MeV$\cdot$fm$^3$&MeV$\cdot$fm$^3$\\
\hline
$^{132}$Ba& 20.0& \cite{bbh85}& 169.2 & 1.42 &0.643& 10.3 &18.0 & 547 & 27.3 & 36.4 \\  
$^{134}$Ba& 20.0& \cite{bbh85}& 130   & 1.42 &0.589& 10.2 &16.0 & 415 & 27.0 & 32.0 \\
$^{136}$Ba& 20.0& \cite{bbh85}& 127.4 & 1.42 &0.608& 10.1 &16.0 & 408 & 26.8 & 31.7 \\
$^{138}$Ba& 20.0& \cite{bbh85}& 128.2 & 1.42 &0.603& 10   &16.0 & 410 & 26.5 & 31.4 \\
$^{140}$Ce& 15.0& \cite{wrt71}& 140   & 1.36 &0.635&  4.4 &23.0 & 398 & 11.7 & 44.6 \\
$^{144}$Sm& 20.0& \cite{mro97}& 100   & 1.42 &0.599&  9.82&14.4 & 319 & 26.0 & 27.4 \\
$^{182}$W & 24.0& \cite{bss81}& 110   & 1.47 &0.556& 13   &15.8 & 382 & 34.2 & 25.2 \\
$^{184}$W & 24.0& \cite{bss81}& 110   & 1.47 &0.550& 13   & 14  & 382 & 34.1 & 22.2 \\
$^{186}$W & 24.0& \cite{bss81}& 110   & 1.47 &0.541& 12.9 &13.6 & 381 & 33.9 & 21.3 \\
$^{192}$Os& 24.0& \cite{bkh76}& 117.4 & 1.47 &0.542& 12.8 &13.2 & 406 & 33.6 & 20.1 \\
$^{197}$Au& 22.0& \cite{rs57} & 110   & 1.44 &0.534& 10.4 &14.0 & 358 & 27.3 & 20.9 \\
	        & 23.65&\cite{kks72}&  95   & 1.46 &0.497& 12.2 &10.0 & 320 & 32.0 & 14.9 \\
	        & 27.95&\cite{kks72}&  90   & 1.48 &0.509& 17   &12.6 & 316 & 44.6 & 18.8 \\
	        & 43.0& \cite{yzr60}&  99.8 & 1.48 &0.508& 33.7 & 0   & 350 & 88.4 &  0 \\
$^{208}$Pb& 23.5& \cite{lfh80}& 110   & 1.46 &0.534& 11.8 &14.0 & 372 & 30.9 & 20.0 \\
	        & 23.6& \cite{kks72}& 110   & 1.46 &0.534& 11.9 &14.0 & 372 & 31.2 & 20.0 \\
	        & 27.6& \cite{kks72}& 103   & 1.48 &0.495& 16.3 &12.0 & 361 & 42.7 & 17.1 \\
$^{209}$Bi& 23.65&\cite{kks72}& 110   & 1.46 &0.500& 11.9 &17.6 & 370 & 31.2 & 25.0 \\
	        & 27.5& \cite{kks72}& 105   & 1.48 &0.503& 16.2 & 4.6 & 368 & 42.4 &  6.5\\
\end{tabular}	 
\end{ruledtabular}
\end{table*}
\endgroup

\begin{figure*} 
\resizebox{2.065\columnwidth}{!}{\includegraphics{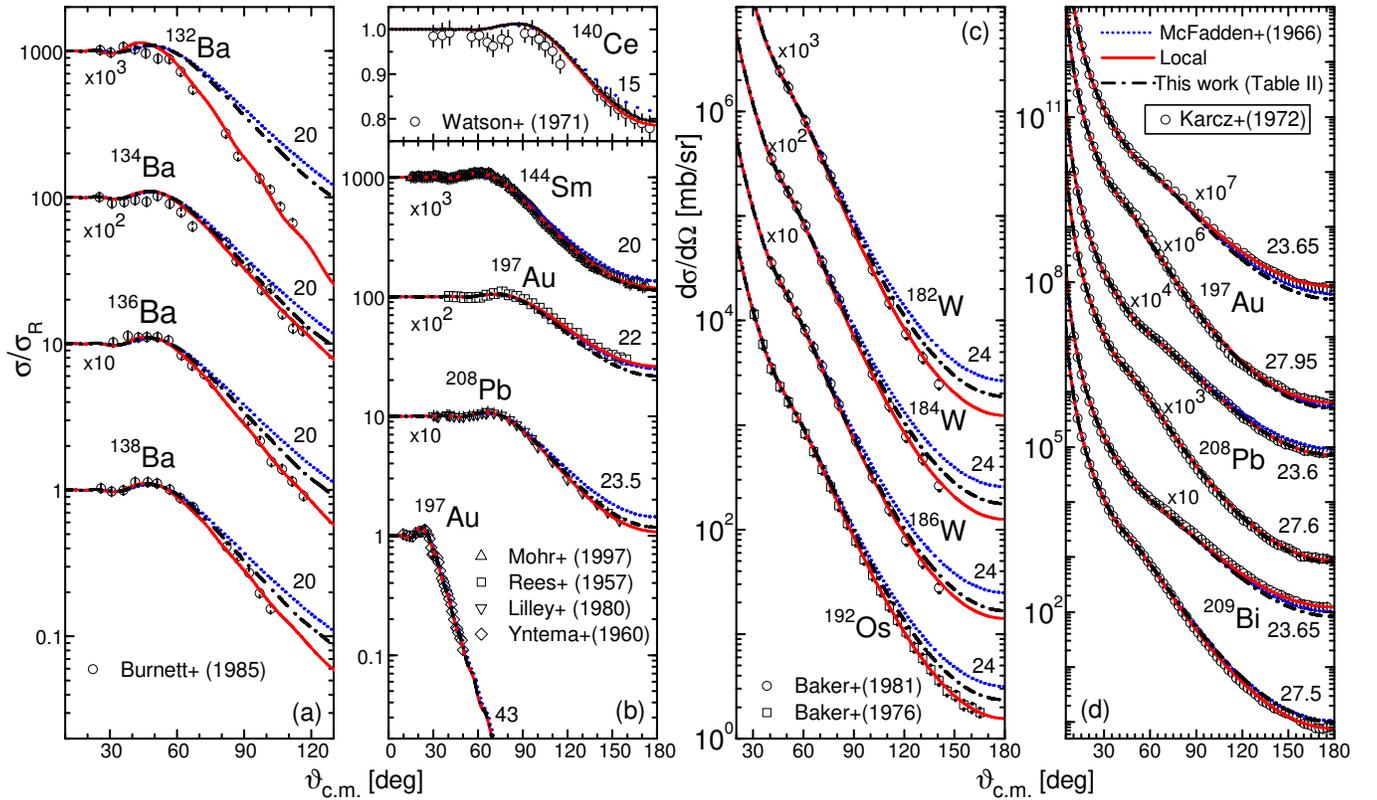}}
\caption{\label{Fig:esad} (Color online) Comparison of experimental angular distributions, either divided by the Rutherford cross section (a),(b) or not (c),(d), of the $\alpha$-particle elastic scattering on the nuclei in Table~\ref{tab:localomp}, and the calculated results corresponding to the local OMP parameters in Table~\ref{tab:localomp} (solid curves) and global sets in Table~\ref{tab:newomp} (dash-dotted curves) and Ref. \cite{mcf66} (dotted curves).}
\end{figure*}

Because the largest sensitivity of the calculated $(\alpha,x)$ reaction data is related to the surface imaginary potential \cite{ma10}, we have considered an increased radius for this potential component. Finally, we found that $\sim$7\% larger values of the $r_D$ parameter may reproduce indeed the experimental $(\alpha,n)$ reaction cross sections for the well-deformed target nuclei (Fig.~\ref{Fig:HoErTmYb}). We have paid the most attention to the $(\alpha,n)$ reaction cross sections owing to their ratios to the corresponding $(\alpha,\gamma)$ reaction cross section with values of already 4--15 at 1.5--2 MeV above their reaction thresholds. This is the case of all target nuclei which form the subject of this work, except the $^{168}$Yb nucleus for which the two reaction cross sections are still almost equal at the above-mentioned energy. While an additional discussion is given for this nucleus in Sec.~\ref{Disc}, we mention here the good agreement obtained also for the $(\alpha,\gamma)$ reaction cross sections of $^{169}$Tm nucleus \cite{ggk11b,tr12}.

\begin{figure} 
\resizebox{0.80\columnwidth}{!}{\includegraphics{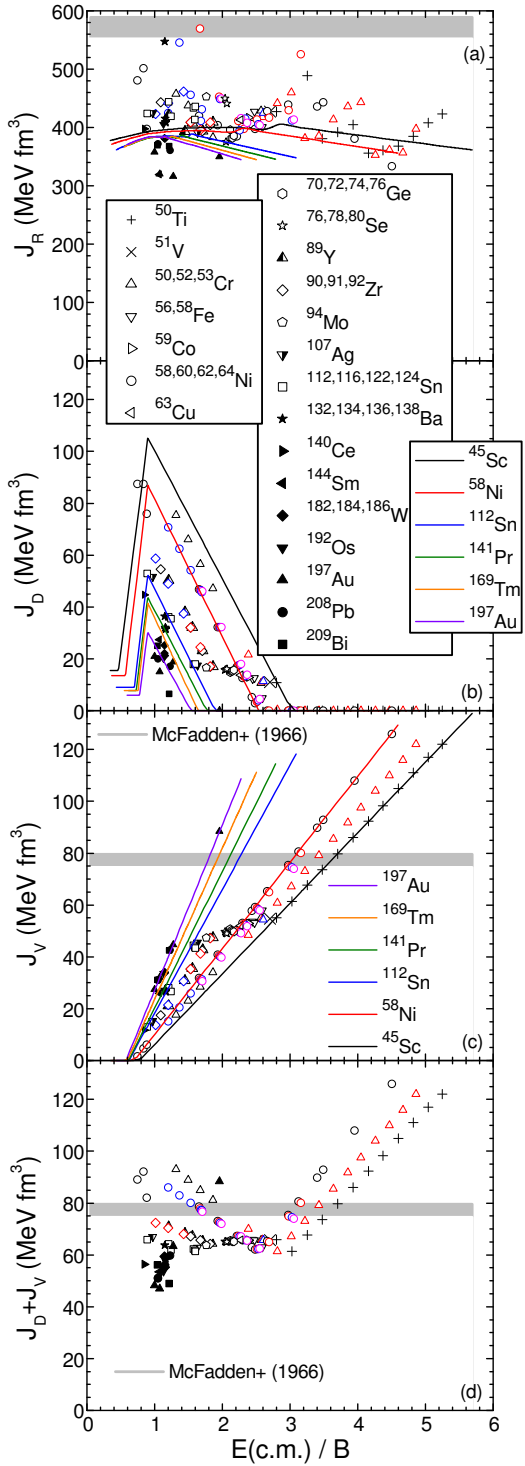}}
\caption{\label{Fig:OPvi} (Color online) Volume integrals per nucleon of the (a) real and (b) imaginary surface and (c) volume components, as well as (d) the sum of the imaginary components, of the local OMP parameter sets obtained through the $\alpha$-particle elastic-scattering analysis for target nuclei within $A$$\sim$50--120 (Table 2 of Ref. \cite{ma09}) and 132--209 \cite{ma09b} (Table~\ref{tab:localomp}), shown by symbols which differ by color for various isotopes of an element. The curves correspond to the OMP in Table~\ref{tab:newomp} for the target nuclei $^{45}$Sc, $^{58}$Ni, $^{112}$Sn, $^{141}$Pr, $^{169}$Tm, and $^{197}$Au, (a),(b) from top to bottom, and (c) in the reverse order, from the lowest incident energies corresponding to the $(\alpha,x)$ reaction data analyzed within Ref. \cite{ma09b} and present work, up to 50 MeV. The shaded regions correspond to the OMP of Ref. \cite{mcf66} for nuclei from $^{45}$Sc to $^{209}$Bi.}
\end{figure}

A similar agreement is obtained for the $(\alpha,\gamma)$ reaction cross sections of $^{162}$Er nucleus \cite{ggk14} but only if, above the $(\alpha,n)$ reaction threshold, a maximum increase with a factor of 2 is obtained by using the SLO model for the $f_{E1}(E_{\gamma})$ $\gamma$-ray strength function. However, this additional change within the SM calculations of this work has no effect on the related $(\alpha,n)$ reaction cross sections and conclusions on the OMP update, owing to a ratio of $\sim$20 between the two reaction cross sections measured at the highest incident energy of Ref. \cite{ggk14}.

\subsection{$\alpha$-particle elastic-scattering account}

The updated surface imaginary potential depth well bellow $B$ has no effect on the description of the $\alpha$-particle elastic-scattering data which have been available  at quite larger energies for target nuclei with $A$$>$130 \cite{bbh85,wrt71,mro97,bss81,bkh76,rs57,kks72,yzr60,lfh80} (Table~\ref{tab:localomp}), formerly analyzed in Ref. \cite{ma09b} within a first step to establish the previous potential \cite{ma10}. The only question mark concerned the case of the $^{182,184,186}$W isotopes, owing to the presently increased surface imaginary potential radius for the well-deformed nuclei. 

Because we have found that the measured $\alpha$-particle elastic-scattering angular distributions at  backward angles are particularly overestimated for these nuclei by the increased $r_D$ value, an energy-dependent form is finally adopted in Table~\ref{tab:newomp} for this potential parameter. Thus, while the $\sim$7\% increased value corresponds to an incident energy of 12 MeV, the global value of 1.52 fm is resumed for the energies of experimental elastic-scattering data, i.e. above $B$. The corresponding uncertainty in  Table~\ref{tab:newomp} also takes into account this yet tentative energy dependence. Unfortunately it could be entirely validated only by further reaction cross-section measurements at low energies. The comparison of the measured and calculated elastic-scattering angular distributions corresponding to the global sets in Table~\ref{tab:newomp} of this work as well as Ref. \cite{mcf66}, and local phenomenological OMP parameters \cite{ma09b} also given in Table~\ref{tab:localomp} are shown in Fig.~\ref{Fig:esad}. 

The volume integrals per nucleon of the real $J_R$ and imaginary surface $J_D$ and volume $J_V$ components of the potential in Table~\ref{tab:newomp}, for several nuclei from $^{45}$Sc to $^{197}$Au, are compared in Fig.~\ref{Fig:OPvi} with the corresponding values of the above-mentioned local phenomenological OMP parameter sets (Table 2 of Ref. \cite{ma09} and Table~\ref{tab:localomp}). The particular curves, given by the average OMP parameters, start from the lowest incident energies corresponding to the $(\alpha,x)$ reaction data analyzed within Ref. \cite{ma09b} and the present work, up to 50 MeV. More details related to these volume integrals per nucleon are given elsewhere \cite{ma03,ma06,ma09}, including the case of microscopic DFM real potentials and the surface and volume imaginary-potential dispersive corrections formerly used within the elastic-scattering analysis. Nevertheless, several additional comments may be given in the following, taking also into account that the volume integrals per nucleon incorporate contributions of both the well depth and its geometry.

The incident energies of the $\alpha$-particle elastic-scattering data which have been available for our analysis, i.e. below 50 MeV, correspond to ratios $E$/$B$ of several units for lighter nuclei but around 1 for heavier ones. At these energies one may observe a decrease of $J_R$, $J_D$, and $J_V$ with the target-nucleus mass, similar to the nucleon case \cite{KD03,jr82}. 

However, the sum of the two imaginary potential components shown in Fig.~\ref{Fig:OPvi}(d) is rather constant for $E$/$B$$\lesssim$3, and close to the sole $J_V$ values corresponding to the OMP of Ref. \cite{mcf66}. Actually, within this energy range our global $J_D$ and  $J_V$ values have a behavior similar to protons below 60 MeV (Figs. 25 and 26 of Ref. \cite{jr82}), i.e., with the $J_D$ first increasing linearly and then decreasing to zero at the same time with the $J_V$ component starting to increase and continue this trend at higher energies. While this behavior corresponds physically to the more inelastic channels which are opened with the incident energy increase, first only within the surface region and then also inside the nucleus, there is a noticeable difference between the protons and $\alpha$-particles OMPs. Thus, in the case of the $\alpha$-particles only the decreasing side of $J_D$ is constrained by the elastic-scattering data while the increasing one could be determined by means of the $\alpha$-induced reaction data analysis. Under the circumstances, the extrapolation to lower energies of an OMP established by analysis of elastic-scattering data may provide better results if it includes only a volume absorption (e.g., the well-known potential of McFadden and Satchler \cite{mcf66}), than in the case that it has also a surface component whose extrapolation would become unphysical (see also Ref. \cite{ma10}).

\section{DISCUSSION} \label{Disc}

\subsection{The $\alpha$-particle induced reactions on $^{168}$Yb}

\begin{figure*} 
\resizebox{1.6\columnwidth}{!}{\includegraphics{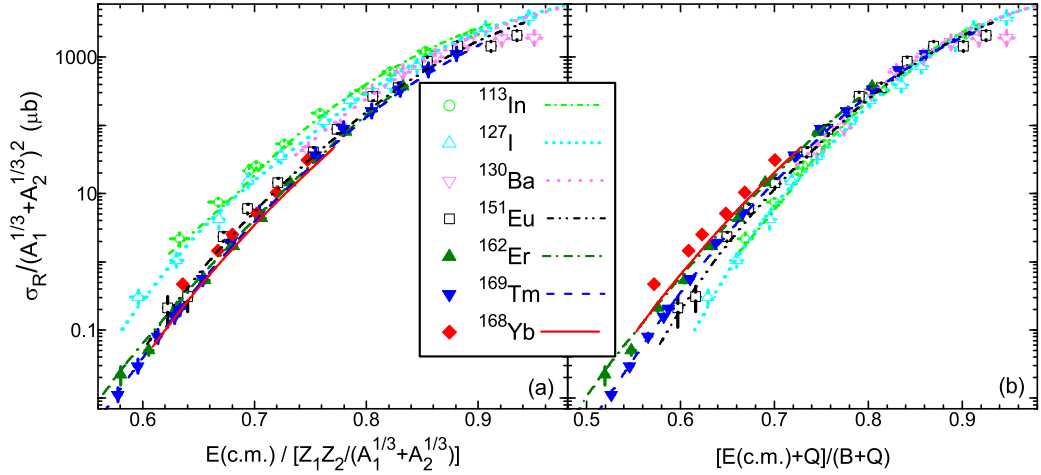}}
\caption{\label{Fig:saRsyst} (Color online) Comparison of the measured and calculated sum of the $(\alpha,n)$ and $(\alpha,\gamma)$ reaction reduced cross sections for the target nuclei $^{113}$In \cite{cy09}, $^{127}$I \cite{ggk12}, $^{130}$Ba \cite{zh12}, $^{151}$Eu \cite{gg10}, $^{162}$Er \cite{ggk14}, $^{169}$Tm \cite{ggk11b,tr12}, and $^{168}$Yb \cite{ln13}, vs reduced energy parameters \cite{prsg04,rw13} (see text).}
\end{figure*}

To understand the possible motivation of the underestimation of measured $(\alpha,\gamma)$ and $(\alpha,n)$ reaction data for the target nucleus $^{168}$Yb \cite{ln13}, we have considered the sum of the $(\alpha,\gamma)$ and $(\alpha,n)$ reaction cross sections which have recently become available for heavier target nuclei, below $B$ \cite{ggk12,zh12,gg10,ggk14,ggk11b,tr12,ln13,cy09}. This sum corresponds actually to the $\alpha$-particle total reaction cross section at the lower energies concerned in the present work while several percents are missing in this respect at the higher energies, owing to the above-mentioned increasing collective inelastic-scattering cross sections. However, for the sake of simplicity and comparison with former systematics, we have taken it into account as $\alpha$-particle $\sigma_R$.

These data, divided by $(A_1^{1/3}+A_2^{1/3})^2$  to eliminate the trivial differences arising from system size, are shown in Fig.~\ref{Fig:saRsyst} versus both (a) the center-of-mass energy divided by $Z_1Z_2/(A_1^{1/3}+A_2^{1/3})$, to eliminate also the differences owing to the barrier height, and (b) the reduced energy parameter proposed recently \cite{rw13} as the ratio between the center-of-mass energy and the Coulomb barrier while to both of them the $Q$ value \cite{ga03} for the CN formation is added. Thus, while the former usual reduction method proposed by Gomes {\it et al.} \cite{prsg04} makes use of a reduced energy which is actually quite close to the ratio $E(c.m.)/B$, the difference in the $Q$ values has also been taken into account for comparison of either  fusion or total reaction cross sections for different systems (e.g., \cite{vm14} and Refs. therein). Nevertheless, both kinds of reduced energies have values $<$1, while the measured reduced cross sections for various target nuclei show an offset from the smooth behavior of total reaction cross sections derived from elastic-scattering data for many $\alpha$-nucleus systems at energies mainly above $B$ \cite{pm11,pm13}. 

While the main aim of this comparison is to highlight the experimental data behavior, there are also shown in Fig.~\ref{Fig:saRsyst} the  calculated cross sections corresponding to the present OMP, including the previous results \cite{ma10} for the target nucleus $^{113}$In. They describe well the above-mentioned offset from a smooth behavior.

Two main comments may follow this comparison. First, the latter reduction method has particularly pointed out larger total reaction cross sections measured for the target nucleus $^{168}$Yb \cite{ln13}, with reference to the closer nuclei $^{162}$Er \cite{ggk14} and $^{169}$Tm \cite{ggk11b,tr12}. Second, the presently calculated results are in agreement with the experimental excitation functions, in the limit of the error bars, except the underestimation of the $^{168}$Yb nucleus data.

The above-mentioned smooth behavior of $\alpha$-particle total reaction cross sections derived from elastic-scattering data for many $\alpha$-nucleus systems, at energies mainly above $B$, was found to be common with the also tightly bound projectile $^{16}$O incident on the semimagic nucleus $^{138}$Ba, unlike weakly bound projectiles \cite{pm11}. However, a rather similar plot but for $^{16}$O$+$$^{144,148,154}$Sm systems has shown quite different sub-barrier fusion enhancement of the corresponding fusion cross sections which match also each other only at energies above $B$ \cite{kh12}. However, this strong target dependence of sub-barrier fusion cross sections was suggested from the beginning by the low-lying spectra of the three Sm isotopes to be attributable to collective excitations of the colliding nuclei during fusion \cite{kh12}. 

While it is already well-known that indeed the Coulomb barrier acts as an amplifier of the couplings associated with a particular degree of freedom, the effects of coupling in the enhancement of the heavy-ion fusion cross sections are lowest for $\alpha$-particles \cite{sg98}. Actually, CC calculations including the internal structure of the colliding nuclei in the dynamics of the reaction have been performed only for heavier-ion induced reactions. Moreover, even if the CC method should be involved for a rigorous treatment of reactions on deformed nuclei, it is considered not suited for large-scale calculations for astrophysics \cite{tr11} and fusion technology as well. Thus, an effective spherical optical potential is yet concerned in this respect,  to obtain SM reliable predictions for nuclei heavier than A$\sim$40 and close to the valley of stability \cite{dr12}. Nevertheless, beyond the present revision of a global OMP, both the involved $\alpha$-particle OMP and the cross-section predictions will benefit from further CC analysis of specific $\alpha$-particle induced reactions.

\begin{figure} [b]
\resizebox{1.0\columnwidth}{!}{\includegraphics{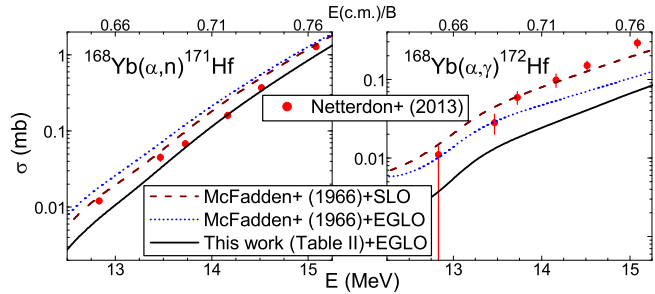}}
\caption{\label{Fig:Yb} (Color online) Comparison of the measured $(\alpha,n)$ and $(\alpha,\gamma)$ reaction cross sections for the target nucleus $^{168}$Yb \cite{ln13} and calculated values using the $\alpha$-particle OMP of Ref. \cite{mcf66} and either the SLO (dashed curves) or EGLO (dotted curves) $\gamma$-ray strength functions, as well as the revised OMP of this work and the EGLO model (solid curves).}
\end{figure}

However, additional calculations were carried out for the $(\alpha,\gamma)$ and $(\alpha,n)$ reaction cross sections for the target nucleus $^{168}$Yb.  Because an eventual agreement was also reported \cite{ln13} by using the $\alpha$-particle OMP of McFadden and Satchler \cite{mcf66} and SLO $f_{E1}(E_{\gamma})$ $\gamma$-ray strength functions, we have looked for the effects of these particular options on the calculated cross sections. An additional note should concern the low-lying levels of the nucleus $^{171}$Hf used within the SM calculations for these reactions. Because the ENSDF evaluation \cite{ensdf} for this nucleus is rather out of date, the latest XUNDL \cite{xundl} unevaluated set was used in this respect (Table~\ref{densp}). This choice leads to a $\sim$40\% increase of the $(\alpha,\gamma)$ reaction cross section at the highest energy of the corresponding new measured data \cite{ln13}.

While the results of SM calculations using EGLO $\gamma$-ray strength functions and both OMPs in Table~\ref{tab:newomp} and Ref. \cite{mcf66} are shown for all reactions discussed in the present work (Figs.~\ref{Fig:TeIBa5},~\ref{Fig:PrSm},~\ref{Fig:HoErTmYb}), we have considered also the SLO model for this target nucleus (Fig.~\ref{Fig:Yb}). First, the use of the latter OMP has already led to larger reaction cross sections, especially at lower incident energies. Then, the replacement of the EGLO $\gamma$-ray strength functions by the SLO ones yields an additional increase that is larger at higher energies owing to the enlarged excitation energies that are thus involved. Therefore, an agreement for the $(\alpha,\gamma)$ reaction cross sections may be obtained at the cost of the use of (i) $\gamma$-ray strength functions at variance with the corresponding measured data, and (ii) the $\alpha$-particle OMP \cite{mcf66} which has been found (e.g., also Refs. \cite{ggk13,jg12,ggk14,tr12,tr11}) to provide usually rather two times larger total reaction cross sections at energies below $B$. 

Nevertheless, the behavior of the $(\alpha,\gamma)$ reaction cross sections for the $^{168}$Yb nucleus, relative to the $(\alpha,n)$ reaction channel, is quite different from that of the other nuclei for which measured data for both reactions exist. We concern in this respect also the rather particular case of the $^{162}$Er nucleus \cite{ggk14} mentioned in Sec.~\ref{OMup}. While the use of SLO $\gamma$-ray strength functions is necessary to describe the measured $(\alpha,\gamma)$ reaction data for both of these nuclei, the ratio of the measured data for the $(\alpha,n)$ and$(\alpha,\gamma)$ reactions is only $\sim$3 for the $^{168}$Yb nucleus at the highest incident energy of Ref. \cite{ln13}, but $\sim$20 for the $^{162}$Er nucleus at the similar energy of Ref. \cite{ggk14}. However, the obvious RSF uncertainties, still present despite the consideration of the corresponding recent achievements, underline the need of more RSF  studies through, e.g., the Oslo method \cite{as00,mg14} or the latest $\beta$-Oslo technique \cite{as14}. Microscopic structure models, either in relativistic or non-relativistic formulation, for the $E1$ and $M1$ RSF. Shell model calculations (e.g., Ref. \cite{bab14}) may also shed light in this respect.

Actually, the particular case of the cross-section ratio for the $^{168}$Yb target nucleus can not be related to the difference between the $Q$-values of the  $(\alpha,\gamma)$ and  $(\alpha,n)$ reaction channels, or the level density of the corresponding residual nuclei, namely the even-even $^{172}$Hf and even-odd $^{171}$Hf for the $\gamma$- and neutron-emission, respectively. 
Therefore, nuclear properties that have not been yet considered may explain the larger $(\alpha,\gamma)$ reaction cross sections for the target nucleus $^{168}$Yb.

\subsection{Updated OMP assessment}

Because the present OMP revision actually concerns only the parameter $W_D$ at energies well below $B$, while the rest of the OMP parameters in Ref. \cite{ma10} are mainly unchanged, the major results and conclusions of Refs. \cite{ma09,ma09b,ma10} are unchanged as well. Concerning these results and conclusions, one may note the excellent fit found by this potential of the recent high precision elastic-scattering data for $^{113}$In at energies near $B$ \cite{ggk13}. There is also a better agreement of calculated results \cite{ma09,ma10} and the $(\alpha,\gamma)$ reaction cross sections for the target nuclei $^{106}$Cd, $^{113}$In and $^{112}$Sn, as well as $(\alpha,n/p)$ reaction cross sections for $^{106}$Cd and  $^{113}$In, in comparison with more recent results of local potentials \cite{ap12,ggk13}. In addition, a recent work on the $^{58}$Ni($\alpha$,$\gamma$)$^{62}$Zn reaction emphasized that further theoretical work is required to obtain a full understanding of $\alpha$-induced reaction cross sections on $^{58}$Ni \cite{sjq14}, i.e. a consistent description of the ($\alpha$,$\gamma$) and $(\alpha,p)$ reaction cross sections together, while this aim was achieved formerly \cite{ma09} for all $^{58,62,64}$Ni isotopes. 

Altogether, this sizable use of the OMP of Ref. \cite{ma10} within significant references motivates the present update which provides improved results and a simpler form too. The latter point is not trivial because the different values of the surface imaginary potential depth and energy limits $E_1$ to be considered for either medium or heavy nuclei, at the lowest $\alpha$-particle energies, could be confusing as they led to unphysical discontinuities of the calculated excitation functions of Ref. \cite{ggk13}. Finally, major question marks with no final decision, even by using various empirical rescaling factors of the $\gamma$ and/or neutron widths, are actual for heavier nuclei \cite{ggk13,jg12,ggk14,tr12} while a consistent and  suitable description of these data is provided within present work.
 
However, the aim of this OMP update is not to answer basic points of the physics of $\alpha$-particle-nucleus interaction at and below $B$. There is indeed a clear progress using latest global \cite{ma10} and folding potentials compared to the older potentials while further improvements of these latest potentials are still required \cite{ggk13}. At the same time it is worthwhile to handle global OMPs within actual computer codes, where it is technically difficult to use, e.g., the folding potentials \cite{pm13}. To help a friendly input, the particular OMP parameters for nuclei involved in Refs. \cite{ma09,ma10} and the present work, as well as tabular forms for the use within the EMPIRE-II \cite{mh02} and TALYS \cite{TALYS} codes are given in the Suppplemental Material \cite{SuppplementalMat}. They should supersede the RIPL-3 subset \cite{ripl3-OMPa-MA} that was based on the previous OMP \cite{ma10}.

\section{CONCLUSIONS} \label{Conc}

The recent high-precision measurements of $\alpha$-particle induced-reaction data below the Coulomb barrier are involved  to understand actual limits and eventually improve the $\alpha$-particle optical-model potential of Ref. \cite{ma10} for nuclei with 45$\leq$$A$$\leq$209, below $B$. Statistical-model calculations of reaction cross sections have been used in this respect, while increased attention has been paid to the use of enhanced forms of SM input parameters that have been obtained in the meantime. Their effects on the calculated $(\alpha,x)$ reaction cross sections have now been found comparable to those given by use of different $\alpha$-particle OMPs. All particular parameter values are given in tabular form or through handy references so that any further analysis could be done right away in similar conditions.

The main revision of the OMP \cite{ma10} concerns actually only one parameter, namely the surface imaginary potential depth at the lowest $\alpha$-particle energies well below $B$, and, in fact, only for the mass range above $A$$\sim$130. Moreover, the updated value is that of the ROP established by analysis of the well-enlarged data basis available for $A$$\sim$50--120 nuclei \cite{ma09}. 

A further regional point has concerned the recent data measured for well-deformed nuclei. The obvious underestimation of both $(\alpha,n)$ and $(\alpha,\gamma)$ reaction cross sections by using the optical potential parameters which have been found suitable for the rest of nuclei is removed if the spherical OMP calculations are performed with $\sim$7\% larger values of the surface imaginary potential radius below the Coulomb barrier. However, beyond the present aim of a revised global OMP, both the involved $\alpha$-particle OMP and the cross-section predictions will benefit from further CC analysis of specific $\alpha$-particle induced reactions. 

Finally, a consistent description is provided for the recent $\alpha$-particle induced reaction data \cite{df11,ap120,ggk12,zh12,as11,gg10,jg12,ggk11b,tr12,ln13}. The only different case is that of the recent data for the $(\alpha,\gamma)$ reaction on $^{168}$Yb \cite{ln13}, which could be partially described by using $\alpha$-particle OMP which provides usually rather two times larger total reaction cross sections at energies below $B$, and $\gamma$-ray strength functions at variance with the corresponding measured data. The need for more RSF studies through the Oslo and $\beta$-Oslo methods is thus pointed out, while shell model calculations as well as further microscopic structure models for the $E1$ and $M1$ RSF would be of largest interest. Nevertheless, further measurements of both $(\alpha,n)$ and $(\alpha,\gamma)$ reaction cross sections for additional target nuclei should increase the actual scarce systematics. The updated $\alpha$-particle global OMP which provides a suitable description of the most $\alpha$-particle induced reaction data could be furthermore involved in a more accurate analysis of the significant underestimation of the $\alpha$-particle emission \cite{ma06,va94}.

\section*{Acknowledgments}

Useful correspondence with Drs. Jura Kopecky and Ionel Stetcu is truly acknowledged. The authors are also indebted to Drs. Jan Glorius,  Anne Sauerwein and Lars Netterdon for making available their data prior to publication. 
This work was partly supported by the Fusion for Energy (F4E-FPA-GRT-168-01), and Ministry of Education, Research, Youth and Sports (www.research.ro/en/) Project No. PN-09-37-01-05.


\begin{thebibliography}{99}
\bibitem{ap12} A. Palumbo {\it et al.}, Phys. Rev. C {\bf 85}, 035808 (2012); {\it ibid.} {\bf 88}, 039902(E) (2013). 
\bibitem{ggk11} G. G. Kiss {\it et al.}, Phys. Rev. C {\bf 83}, 065807 (2011).
\bibitem{ggk13} G. G. Kiss {\it et al.}, Phys. Rev. C {\bf 88}, 045804 (2013).
\bibitem{df11} D. Filipescu {\it et al.}, Phys. Rev. C {\bf 83}, 064609 (2011).
\bibitem{ap120} A. Palumbo, W. P. Tan, J. G\"orres, M. Wiescher, N. \"Ozkan, R. T. G\"uray, and C. Yal\c cin, Phys. Rev. C {\bf 85}, 028801 (2012).
\bibitem{ggk12} G. G. Kiss, T. Sz\"ucs, Zs. T\"or\"ok, Z. Korkulu, Gy. Gy\"urky, Z. Hal\'asz, Zs. F\"ul\"op, E. Somorjai, and T. Rauscher, Phys. Rev. C {\bf 86}, 035801 (2012).
\bibitem{zh12} Z. Hal\'asz, Gy. Gy\"urky, J. Farkas, Zs. F\"ul\"op, T. Sz\"ucs, E. Somorjai, and T. Rauscher, Phys. Rev. C {\bf 85}, 025804 (2012).
\bibitem{as11} A. Sauerwein {\it et al.}, Phys. Rev. C {\bf 84}, 045808 (2011).
\bibitem{gg10} Gy. Gy\"urky {\it et al.}, J. Phys. G {\bf 37}, 115201 (2010).
\bibitem{jg12} J. Glorius, J. G\"orres, M. Kn\"orzer, R. Reifarth, A. Sauerwein, K. Sonnabend, and M. Wiescher, in: {\it XII International Symposium on Nuclei in the Cosmos, August 5-12, 2012, Cairns, Australia}, PoS(NIC XII)118;  http://pos.sissa.it/cgi-bin/reader/conf.cgi?confid=146; J. Glorius {\it et al.}, Phys. Rev. C {\bf 89}, 065808 (2014).
\bibitem{ggk14}  G.G. Kiss, T. Sz\"ucs, T. Rauscher, Zs. T\"or\"ok, Zs. F\"ul\"op, Gy. Gy\"urky, Z. Hal\'asz, and E. Somorjai, Phys. Lett. B {\bf 735}, 40 (2014). 
\bibitem{ggk11b}  G.G. Kiss, T. Rauscher, T. Sz\"ucs, Zs. Kert\'esz, Zs. F\"ul\"op, Gy. Gy\"urky, C. Fr\"ohlich, J. Farkas, Z. Elekes, and E. Somorjai, Phys. Lett. B {\bf 695}, 419 (2011); G.G. Kiss, T. Sz\"ucs, Gy. Gy\"urky, Zs. F\"ul\"op, J. Farkas, Zs. Kert\'esz, E. Somorjai, M. Laubenstein, C. Fr\"ohlich, and T. Rauscher, Nucl. Phys. {\bf A867}, 52 (2011).
\bibitem{tr12} T. Rauscher, G.G. Kiss, T. Sz\"ucs, Zs. F\"ul\"op, C. Fr\"ohlich, Gy. Gy\"urky, Z. Hal\'asz, Zs. Kert\'esz, and E. Somorjai, Phys. Rev. C {\bf 86}, 015804 (2012).
\bibitem{ln13} L. Netterdon, P. Demetriou, J. Endres, U. Giesen, G. G. Kiss, A. Sauerwein, T. Sz\"ucs, K. O. Zell, and A. Zilges, Nucl. Phys. {\bf A916}, 149 (2013).

\bibitem{ma09} M. Avrigeanu, A. C. Obreja, F. L. Roman, V. Avrigeanu, and W. von Oertzen, At. Data Nucl. Data Tables {\bf 95}, 501 (2009).
\bibitem{ma10} M. Avrigeanu and V. Avrigeanu, Phys. Rev. C {\bf 82}, 014606 (2010).
\bibitem{va11} V. Avrigeanu and M. Avrigeanu, Phys. Rev. C {\bf 83}, 017601 (2011). 

\bibitem{pm11} P. Mohr, Phys. Rev. C {\bf 84}, 055803 (2011).
\bibitem{pm13} P. Mohr, Phys. Rev. C {\bf 87}, 035802 (2013).
\bibitem{tr11} T. Rauscher, Int. J. Mod. Phys. E {\bf 20}, 1071 (2011).
\bibitem{sjq14} S. J. Quinn {\it et al.}, Phys. Rev. C {\bf 89}, 054611 (2014).

\bibitem{ma06} M. Avrigeanu,  W. von Oertzen, and V. Avrigeanu, Nucl. Phys. {\bf A764}, 246 (2006).
\bibitem{go02} P. Demetriou, C. Grama, and S. Goriely,  Nucl. Phys. {\bf A707}, 253 (2002).
\bibitem{ma03} M. Avrigeanu, W. von Oertzen, A.J.M. Plompen, and V. Avrigeanu, Nucl. Phys. {\bf A723}, 104 (2003). 
\bibitem{ma10b} M. Avrigeanu and V. Avrigeanu, Phys. Rev. C {\bf 81}, 038801 (2010). 
\bibitem{ma09b} M. Avrigeanu, W. von Oertzen, A. C. Obreja, F. L. Roman, and V. Avrigeanu, in {\it Proc. 12th Int. Conf. on Nuclear Reaction Mechanisms, 15-19 June 2009, Villa Monastero, Varenna, Italy}, edited by F. Cerutti and A. Ferrari, CERN-Proceedings-2010-001-V-1 (CERN, Geneva, 2010); http://cdsweb.cern.ch/record/1237907/files/p159.pdf.

\bibitem{ND2013} V. Avrigeanu and M.~Avrigeanu, in {\it Proceedings of the International Conference on Nuclear Data for Science and Technology, New York, March 2013}, Nucl. Data Sheets {\bf 118}, 262 (2014).
\bibitem{cssp14} V. Avrigeanu, M.~Avrigeanu,  and C. M\u an\u ailescu, in {\it Exotic Nuclei and Nuclear/Particle Astrophysics (V). From Nuclei to Stars: Carpathian Summer School of Physics 2014, Sinaia, Romania}, AIP Conf. Proc. (to be published); http://cssp14.nipne.ro/
\bibitem{tr12b} T. Rauscher, in: {\it XII International Symposium on Nuclei in the Cosmos, August 5-12, 2012, Cairns, Australia}, PoS(NIC XII)052;  http://pos.sissa.it/cgi-bin/reader/conf.cgi?confid=146 
\bibitem{tr13} T. Rauscher, Phys. Rev. Lett. {\bf 111}, 061104 (2013).
\bibitem{tr01} T. Rauscher and F.-K. Thielemann, At. Data Nucl. Data Tables {\bf 79}, 47 (2001).
\bibitem{scat2} O.~Bersillon, Code SCAT2, Note CEA-N-2227, 1992.
\bibitem{mcf66} L. McFadden and G.R. Satchler, Nucl. Phys. {\bf A84}, 177 (1966).
\bibitem{es98} E. Somorjai {\it et al.}, Astron. Astrophys. {\bf 333}, 1112 (1998).
\bibitem{hv83} H. Vonach, R. C. Haight, and G. Winkler, Phys. Rev. C {\bf 28}, 2278 (1983).
\bibitem{msh91} M. S. Hussein, R. A. Rego, and C. A. Bertulani, Phys. Rep. {\bf 201}, 279 (1991), p. 285.
\bibitem{wn80} W. N\"orenberg, in {\it Heavy Ion Collisins}, Vol. 2, Edited by R. Bock (North-Holland, Amsterdam, 1980), p. 11.
\bibitem{KD03} A. J. Koning and J. P. Delaroche, Nucl. Phys. {\bf A713}, 231 (2003).
\bibitem{hv88} H. Vonach, M. Uhl, B. Strohmaier, B. W. Smith, E. G. Bilpuch, and G. E. Mitchell, Phys. Rev. C {\bf 38}, 2541 (1988).
\bibitem{exfor}Experimental Nuclear Reaction Data (EXFOR), http://www-nds.iaea.org/exfor/ 
\bibitem{ensdf} Evaluated Nuclear Structure Data File (ENSDF), http://www.nndc.bnl.gov/ensdf/ 
\bibitem{xundl} Experimental Unevaluated Data (XUNDL), http://www.nndc.bnl.gov/ensdf/ensdf/xundl.jsp  
\bibitem{ripl3} R. Capote {\it et al.}, Nucl. Data Sheets {\bf 110}, 3107 (2009); http://www-nds.iaea.org/RIPL-3/
\bibitem{ma95} M. Avrigeanu and V. Avrigeanu, IPNE Report NP-86-1995, Bucharest, 1995, and Refs. therein; News NEA Data Bank {\bf 17}, 22 (1995).
\bibitem{va02} V. Avrigeanu, T. Glodariu, A. J. M. Plompen, and H. Weigmann, J. Nucl. Sci. Technol. Suppl. {\bf 2}, 746 (2002); http://tandem.nipne.ro/\verb|~|vavrig/publications/2002/Tables/caption.html

\bibitem{mg14}http://www.mn.uio.no/fysikk/english/research/about/ infrastructure/OCL/nuclear-physics-research/compilation/ 
\bibitem{ripl2} T. Belgya {\it et al.}, Tech. Rep. IAEA-TECDOC-1506, IAEA, Vienna, Austria, 2006, p. 97; https://www-nds.iaea.org/RIPL-2/gamma/gamma-strength-exp.dat

\bibitem{as00} A. Schiller, L. Bergholt, M. Gutteormsen, E. Melby, J. Rekstad, ans S. Siem, Nucl. Instr. Meth. A {\bf 447}, 498 (2000).
\bibitem{dgg79} D. G. Gardner and F. S. Dietrich, Report Lawrence	Livermore National Laboratory UCRL-82998, Livermore, 1979.
\bibitem{ma87} M. Avrigeanu, V. Avrigeanu, G. C\u ata, and 	M. Ivascu, Rev. Roum. Phys. {\bf 32}, 837 (1987).
\bibitem{jk91} J. Kopecky and M. Uhl, Phys. Rev. C {\bf 41}, 1941 (1990).
\bibitem{jk93} J. Kopecky, M. Uhl, and R. E. Chrien, Phys. Rev. C {\bf 47}, 312 (1993).
\bibitem{ripl1} P. Oblo\v zinsk\' y {\it et al.}, Tech. Rep. IAEA-TECDOC-1034, IAEA, Vienna, Austria, 1998; http://www-nds.iaea.org/ripl/ 
\bibitem{bb13} B. Baramsai {\it et al.}, Phys. Rev. C {\bf 87}, 044609 (2013); J. Kroll {\it et al.}, Phys. Rev. C {\bf 88}, 034317 (2013).
\bibitem{acl10} A. C. Larsen and S. Goriely, Phys. Rev. C {\bf 82}, 014318 (2010).
\bibitem{hkt11} H. K. Toft {\it et al.}, Phys. Rev. C {\bf 83}, 044320 (2011); A.C. Larsen {\it et al.}, Phys. Rev. C {\bf 87}, 014319 (2013).
\bibitem{pdk84}P.D. Kunz, DWUCK4 user manual, OECD/NEA Data Bank,
	Issy-les-Moulineaux, France, 1984; http://www.oecd-nea.org/tools/abstract/detail/nesc9872
\bibitem{pgy86} P. G. Young, Report NEANDC-222"U", OECD/NEA Data Bank, Gif-sur-Yvette, 1986, p. 127.
\bibitem{gn09} G. Noguere, E. Rich, C. De Saint Jean, O. Litaize, P. Siegler, and V. Avrigeanu, Nucl. Phys. {\bf A831}, 106 (2009).
\bibitem{mh02} M. Herman, EMPIRE-II v.2.18 dated 2002-09-27; http://www-nds.iaea.org/empire/ 
\bibitem{kw06} K. Wisshak, F. Voss, F. K\" appeler, L. Kazakov, F. Be\v cv\' ar, M. Krti\v cka, R. Gallino, and M. Pignatari, Phys. Rev. C {\bf 73}, 045807 (2006).
\bibitem{rgl90} R. G. Lanier, H. I. West, Jr., M. G. Mustafa, J. Frehaut, A. Adam, and C. A. Philis, Phys. Rev. C {\bf 42}, R479 (1990).
\bibitem{smg13} S. M. Grimes, Phys. Rev. C {\bf 88}, 024613 (2013).

\bibitem{bbh85} S. M. Burnett, A. M. Baxter, S. Hinds, F. Pribac, R. Smith, R. H. Spear, and M. P. Fewell, Nucl. Phys. {\bf A442}, 289 (1985); EXFOR-F0829 data file entry.
\bibitem{wrt71} B. D. Watson, D. Robson, D. D. Tolbert, and R. H. Davis, 	Phys. Rev. C {\bf 4}, 2240 (1971); EXFOR-C1386 data file entry.
\bibitem{mro97} P. Mohr, T. Rauscher, H. Oberhummer, Z. M\'at\'e, Zs. F\"ul\"op,	E. Somorjai, M. Jaeger, and G. Staudt, Phys. Rev. C {\bf 55}, 1523 (1997). 
\bibitem{bss81} F. T. Baker, A. Scott, R. C. Styles, T. H. Kruse,	K. Jones, and R. Suchannek, Nucl. Phys. {\bf A351}, 63 (1981); EXFOR-C0890	data file entry.
\bibitem{bkh76} F. T. Baker, T. H. Kruse, W. Hartwig, I.Y. Lee, and J. X. Saladin, Nucl. Phys. {\bf A258}, 43 (1976); EXFOR-C0894 data file entry.
\bibitem{rs57} J. R. Rees and M. B. Sampson, Phys. Rev. {\bf 108},1289 (1957) ; EXFOR-C1054 data file entry.
\bibitem{kks72} W. Karcz, I. Kluska, Z. Sanok, J. Szmider, J. Szymakowski, S. Wiktor, and R. Wolski, Acta Physica Polonica B {\bf 3}, 525 (1972); EXFOR-F0700 data file entry. 
\bibitem{yzr60} J. L. Yntema, B. Zeidman, and B. J. Raz, Phys. Rev. {\bf 117}, 801 (1960); EXFOR-C1082 data file entry.
\bibitem{lfh80} J. S. Lilley, M. A. Franey, and Da Hsuan Feng, Nucl. Phys. {\bf A342}, 165 (1980); EXFOR-C1282 data file entry.
\bibitem{jr82} J. Rapaport, Phys. Rep. {\bf 87}, 25 (1982).

\bibitem{cy09} C. Yal\c cin {\it et al.}, Phys. Rev. C {\bf 79},  065801 (2009).
\bibitem{rw13} R. Wolski, Phys. Rev. C {\bf 88}, 041603(R) (2013).
\bibitem{ga03} G. Audi, A. H. Wapstra, and C. Thibault, Nucl. Phys. {\bf A729}, 337 (2003); http://www.nndc.bnl.gov/qcalc/
\bibitem{prsg04} P. R. S. Gomes {\it et al.}, Phys. Lett. B {\bf 601}, 20 (2004).
\bibitem{vm14} V. Morcelle {\it et al.}, Phys. Rev. C {\bf 89}, 044611 (2014).
\bibitem{kh12} K. Hagino and N. Takigawa, Prog. Theor. Phys. {\bf 128}, 1061 (2012).
\bibitem{sg98}S. Gil and D. E. DiGregorio, Phys. Rev. C {\bf 57}, R2826 (1998).
\bibitem{dr12} D. Robertson, J. G\"orres, P. Collon, M. Wiescher, and H.-W. Becker, Phys. Rev. C {\bf 85}, 045810 (2012).
\bibitem{as14} A. Spyrou {\it et al.}, http://arxiv.org/abs/1408.6498
\bibitem{bab14} B. A. Brown and A. C. Larsen, http://arxiv.org/abs/1409.3492
\bibitem{TALYS} A. J. Koning, S.~Hilaire, and M. C. Duijvestijn, v. TALYS-1.6, 2013, http://www.talys.eu
\bibitem{SuppplementalMat} See Supplemental Material at http://link.aps.org/sup-plemental/10.1103/PhysRevC.xx.xxxxxx for OMP parameters for nuclei involved in Refs. [15,16] and the present work, as well as tabular forms for the use within the EMPIRE-II [60] andTALYS [84] codes.
\bibitem{ripl3-OMPa-MA} https://www-nds.iaea.org/RIPL-3/optical/om-parameter-u.dat , ‘iref’=9603-9678.
\bibitem{va94} V. Avrigeanu, P.E. Hodgson, and M. Avrigeanu, Phys. Rev. C {\bf 49}, 2136 (1994).

\end{thebibliography}
\end{document}